\documentclass[12pt,a4paper]{article}

\usepackage{amsmath}
\usepackage{amssymb}
\usepackage{hyperref}
\usepackage{dsfont}
\usepackage{hyperref}
\usepackage{graphics}
\usepackage{graphicx}
\usepackage{epsfig}

\let\oldappendix=\appendix
\let\oldsection=\section
\renewcommand{\appendix}{\oldappendix%
\def\theequation{\Alph{section}.\arabic{equation}}%
\renewcommand{\section}{\setcounter{equation}{0}\oldsection}}

\newcommand{\beq}{\begin{equation}}
\newcommand{\eeq}{\end{equation}}
\newcommand{\beqa}{\begin{eqnarray}}
\newcommand{\eeqa}{\end{eqnarray}}
\newcommand{\no}{\nonumber}

\newcommand{\qq}{\qquad}

\newcommand{\newop}[2]{\def#1{\mathop{\mathrm{#2}}\nolimits}}
\newop{\artanh}{artanh}

\newcommand{\Lagr}{\mathcal{L}}

\begin{document}

\hfill 

\hfill 

\bigskip\bigskip

\begin{center}

{{\Large\bf Can one see the number of colors 
          in $\mbox{\boldmath$\eta, \eta' \rightarrow \pi^+ \pi^- \gamma$}$? }}
\vspace{.4in}

{\large B. Borasoy$^{a,b}$ \footnote{email: borasoy@itkp.uni-bonn.de} and 
E. Lipartia$^{a}$ \footnote{email: lipartia@ph.tum.de}}

\bigskip

\bigskip

$^a$ Physik Department,
Technische Universit{\"a}t M{\"u}nchen,\\
85747 Garching, Germany \\[0.4cm]
$^b$ Helmholtz-Institut f\"ur Strahlen- und Kernphysik (Theorie),
Universit{\"a}t Bonn,\\
Nu{\ss}allee 14-16, 53115 Bonn, Germany \\[0.4cm]

\vspace{.2in}

\end{center}

\vspace{.7in}

\thispagestyle{empty} 

\begin{abstract}
We investigate the decays $\eta, \eta' \rightarrow \pi^+ \pi^- \gamma$
up to next-to-leading order in the framework
of the combined $1/N_c$ and chiral expansions. 
Counter terms of unnatural parity at next-to-leading order with unknown 
couplings are important to acommodate the results both to the experimental 
decay width and the photon spectrum. The presence of these coefficients 
does not allow for a determination of the number of colors from these decays.
\end{abstract}\bigskip

\begin{center}
\begin{tabular}{ll}
\textbf{PACS:}&12.39.Fe\\[6pt]
\textbf{Keywords:}& Chiral Lagrangians, large $N_c$, chiral anomalies.
\end{tabular}
\end{center}

% 11.30.Rd Chiral symmetries
% 12.39.Fe Chiral Lagrangians
% 12.40.Vv Vector-meson dominance
% 13.20.Cz Decays of pi mesons
% 13.40.Hq Electromagnetic decays

\vfill

\newpage

The anomalous decay $\pi^0 \to \gamma \gamma$ is presented as a textbook
example to confirm from low-energy hadron dynamics the number of colors to be
$N_c=3$, see e.g. \cite{DGH}, since
this decay originates at tree level from the Wess-Zumino-Witten (WZW) term
 \cite{WZ,W}
with a quantized prefactor $N_c$. The decay width 
$\Gamma_{\pi^0 \rightarrow \gamma \gamma}$ is thus
proportional to $N_c^2$, being quite sensitive to the number of colors, and 
in fact the result for $N_c=3$ is in perfect agreement with experiment.

Recently, however, it was shown 
in \cite{AB,BW} that the cancellation of triangle anomalies in the standard 
model with an arbitrary number of colors leads to $N_c$ dependent values of 
the quark charges, such that the vertex with one pion
and two photons is completely canceled by the $N_c$ dependent part of a
Goldstone-Wilczek term \cite{BW, GW}. 
Within this scenario the decay $\pi^0 \to \gamma \gamma$ cannot be utilized
to support $N_c =3$.
A similar cancellation also occurs for the
decay $\eta \rightarrow \gamma \gamma$, if one neglects $\eta$-$\eta'$ mixing.
The $N_c$ independence is maintained at one-loop order, i.e. at 
next-to-next-to-leading
order in the combined chiral/large $N_c$ expansion, for both the $\pi^0$ and 
the $\eta$ decay,
but the strong $N_c$ dependence of the
singlet decay $\eta_0 \to \gamma \gamma$ induces also a strong $N_c$ 
dependence 
for $\eta \rightarrow \gamma \gamma$ due to $\eta$-$\eta'$ mixing \cite{B}.
One concludes then that both the $\eta$ and the $\eta'$ decay show clear
evidence that we live in a
world with three colors.

On the other hand, it has been pointed out in \cite{BW} that at tree level the 
decay width of the process
$\eta \to \pi^+ \pi^- \gamma$ is proportional to $N_c^2$ and should replace the
textbook process $\pi^0 \to \gamma \gamma$ lending support to $N_c=3$.
In analogy to the two-photon decays, the effects of $\eta$-$\eta'$ mixing
along with the inclusion of subleading contributions must be treated
systematically, in order
to make a rigorous statement on a possible determination of the number 
of colors
from this process.
In the present work we will therefore investigate 
the decays $\eta, \eta' \rightarrow \pi^+ \pi^- \gamma$
up to next-to-leading order
within the framework of large $N_c$ chiral perturbation theory (ChPT) 
\cite{KL1}.

At leading order in the combined chiral and $1/N_c$ expansions the decays
$\eta, \eta' \rightarrow \pi^+ \pi^- \gamma$ originate from a piece in the
WZW Lagrangian
\beq  \label{wzwvec}
S_{\scriptscriptstyle{WZW}} (U, v) = - \frac{N_c}{48 \pi^2} \int \; 
\langle \Sigma_{\scriptscriptstyle{L}}^3 v -  
\Sigma_{\scriptscriptstyle{R}}^3 v \rangle ,
\eeq
where $\Sigma_{\scriptscriptstyle{L}} = U^\dagger dU $, 
$\Sigma_{\scriptscriptstyle{R}} = U dU^\dagger $, 
and we adopted the differential form notation of \cite{KL1},
\beq
v = dx^\mu v_\mu ,\qq d = dx^\mu \partial_\mu 
\eeq
with the Grassmann variables $dx^\mu$ which yield the volume element $dx^\mu
dx^\nu dx^\alpha dx^\beta = \epsilon^{\mu \nu \alpha \beta} d^4x$.
The brackets $\langle \ldots \rangle$ denote the trace in flavor space,
while the unitary matrix $U= e^{i \phi}$ collects the pseudoscalar meson nonet 
$(\pi, K, \eta_8, \eta_0)$. 
The external vector field $v = - e Q A$ contains the photon field 
$A= A_\mu dx^\mu $
and the quark charge matrix $Q$ of the $u$- $d$- and $s$-quarks which is 
usually
assumed to be independent of the number of colors with $Q = \frac{1}{3} 
\mbox{diag} (2, -1,-1)$.
However, the cancellation of triangle anomalies requires $Q$ to depend 
on $N_c$ \cite{AB,BW}
\beqa
Q &=& \frac{1}{2}  \mbox{diag} \Big( \frac{1}{N_c} +1,\frac{1}{N_c} -1,
\frac{1}{N_c} -1 \Big)\no\\%[2mm]
&=&  \hat{Q} + \Big( 1 - \frac{N_c}{3}  \Big)  \frac{1}{2 N_c} \mathds{1} 
\eeqa
with $\hat{Q} = \frac{1}{3} \mbox{diag} (2, -1,-1)$ being the conventional
charge matrix,
while the second term is proportional to the baryon number and gives rise 
to the
Goldstone-Wilczek term.
The anomalous Lagrangian of Eq.~(\ref{wzwvec}) decomposes into the conventional
WZW Lagrangian of the $U(3)$ theory with the charge matrix $\hat{Q}$ and a 
Goldstone-Wilczek term
which vanishes for $N_c =3$
\beq
S_{\scriptscriptstyle{WZW}} (U, v) = S_{\scriptscriptstyle{WZW}} (U, \hat{v})
+  \Big( 1 - \frac{N_c}{3}  \Big)  S_{\scriptscriptstyle{GW}} (U, A)
\eeq
with $\hat{v} = -e \hat{Q} A$ and
\beqa
S_{\scriptscriptstyle{WZW}} (U, \hat{v}) &=& \frac{ N_c e}{48 \pi^2} \int \;
   \big\langle \, (\Sigma_{\scriptscriptstyle{L}}^3 
   - \Sigma_{\scriptscriptstyle{R}}^3)\,  \hat{Q} \, \big\rangle \, A ,  
\no \\%[0.3cm]
 S_{\scriptscriptstyle{GW}} (U, A)  &=& \frac{ e}{48 \pi^2} \int \;
   \big\langle \, \Sigma_{\scriptscriptstyle{L}}^3 \, \big\rangle \, A  .
\eeqa
However, this presentation is not convenient to perform calculations within
the framework of large $N_c$ ChPT. To this end, one rather 
expands the quark charge matrix $Q$ in powers of  $1/N_c$
\beqa
Q  =  \frac{1}{2}  \mbox{diag} ( 1,-1,-1) + \frac{1}{2 N_c} \mathds{1} 
  \equiv   Q^{(0)}  + Q^{(1)} , 
\eeqa
where the superscript denotes the order in the combined large $N_c$ and chiral 
counting scheme,
i.e. $Q^{(0)}$ ($Q^{(1)}$) is of order ${\cal O}(1)$ (${\cal O}(\delta)$).
From $S_{\scriptscriptstyle{WZW}}$ one obtains the tree level contributions 
\beqa  \label{wzwvecexp}
S_{\scriptscriptstyle{WZW}} (U, v) &=& \int \; d^4 x \; 
{\cal L}_{\scriptscriptstyle{WZW}} 
= - \frac{i N_c e}{24 \pi^2} \int \; 
\langle d \phi \, d \phi \, d \phi \, Q \rangle A \no \\%[2mm]
&=& - \frac{i N_c e}{24 \pi^2} \int \; 
\langle d \phi \, d \phi \, d \phi \, Q^{(0)} \rangle A ,
\eeqa
since for the processes $\eta_8, \eta_0 \rightarrow \pi^+ \pi^- \gamma$
the trace with $Q^{(1)}$ in Eq.~(\ref{wzwvecexp}) vanishes.
The pertinent amplitudes have the structure
\beq  \label{treeamp}
{\cal A}^{(tree)}(\phi \to  \pi^+ \pi^- \gamma)= - \frac{N_c e}{\sqrt{3} \, 
12 \pi^2 f^3 }
       k_\mu \epsilon_\nu p^+_\alpha p^-_\beta \epsilon^{\mu \nu \alpha \beta} 
\alpha^{(tree)}_\phi ,
\eeq
where $p^+ (p^-)$ is the momentum of the outgoing $\pi^+ (\pi^-)$ and
$k (\epsilon)$ is the momentum (polarization) of the outgoing photon.
Next we replace $f^3$ by $F_\phi F_\pi^2$ in Eq.~(\ref{treeamp}) with
the decay constants $F_\varphi $ defined via
\beq  \label{decconst}
\langle 0 | \bar{q} \gamma_\mu \gamma_5  \lambda^i  
q |\varphi \rangle = i \sqrt{2} \; p_\mu F_\varphi^i 
\eeq
which is consistent at leading order.
Neglecting $\eta$-$\eta'$ mixing for the moment, 
the coefficients $\alpha^{(tree)}_\phi$ read 
\beq
\alpha^{(tree)}_\eta = 1 \; , \qquad \alpha^{(tree)}_{\eta'} = \sqrt{2} \; .
\eeq
These are the expressions which were suggested to be utilized for 
a determination
of $N_c$ \cite{BW} 
\footnote{Note, however, that a factor of 1/3 is missing in
  the 
amplitudes given 
in \cite{BW}.}.
Employing the experimental values \cite{PDG}
\beqa\label{exprvalues}
\Gamma_{\eta \to \pi^+ \pi^- \gamma} &=&  56.1 \pm 5.4 \; \mbox{eV} , \no \\
\Gamma_{\eta' \to \pi^+ \pi^- \gamma} &=&  59.6 \pm 5.2 \; \mbox{keV} ,
\eeqa
we extract from the $\eta$ decay $N_c =7$ 
and $N_c =10$ from the
$\eta'$ decay which is clearly in contradiction to the well-established
value $N_c=3$.

Taking into account $\eta$-$\eta'$ mixing at leading order
\beqa  \label{eq:leadmix}
\eta_8 &=& \cos \vartheta^{(0)} \; \eta  + \sin \vartheta^{(0)} \; \eta' \no \\
\eta_0 &=&  - \sin  \vartheta^{(0)} \; \eta + \cos \vartheta^{(0)} \;  \eta' 
\eeqa
with the mixing angle $\vartheta^{(0)}$ given by
\beq
\sin 2 \vartheta^{(0)} = - \frac{4 \sqrt{2}}{3} 
\frac{m_K^2 - m_\pi^2}{m_{\eta'}^2 - m_\eta^2} 
\eeq
the experimental values given in Eq.~(\ref{exprvalues}) allow either for 
$N_c =4$ or $N_c =5$, but $N_c=3$ is clearly ruled out. 
We can therefore conclude that the decays
$\eta, \eta' \rightarrow \pi^+ \pi^- \gamma$ at leading order
are not suited to confirm the number of colors.
In the following we investigate whether the situation changes
by including next-to-leading order corrections.

At next-to-leading order gauge invariant counter terms of unnatural
parity enter the calculation.
First, there is a term of fourth chiral order which is suppressed
by one order in $N_c$ with respect to the leading order result 
\cite{KL1, BN1, BN2}
\beq
d^4x \, {\tilde \Lagr}_{p^4}=i {\tilde L}_1 \psi \langle dv \, dU dU^{\dagger} 
+ dv \, dU^{\dagger} dU\rangle
\eeq
with $\psi = -i  \ln \det U$ and we have neglected for brevity both the 
external axial-vector fields and the QCD vacuum angle $\theta$. 

At the same order in the $\delta$ expansion of large $N_c$ ChPT counter terms
of sixth chiral order contribute which can be decomposed into explicitly
symmetry breaking terms and terms with additional derivatives \cite{BN2,ChPTO6}
\beq
{\tilde \Lagr}_{p^6} = {\tilde \Lagr}_{\chi} + {\tilde \Lagr}_{\partial^2} ,
\eeq
where
\beqa
d^4x \, {\tilde \Lagr}_{\chi}&=&{\tilde K}_1\langle\left(U^{\dagger}\chi - 
\chi^{\dagger}U\right)\left[\left(U^{\dagger} dv U + dv\right) U^{\dagger}
dU U^{\dagger}dU 
+U^{\dagger}dU U^{\dagger} dU\left(U^{\dagger} dv U +
 dv\right)\right]\rangle\nonumber\\%[2mm]
&+&  {\tilde K}_2\langle\left(U^{\dagger}\chi - \chi^{\dagger}U\right)
U^{\dagger} dU \left(U^{\dagger} dv U + dv \right)U^{\dagger} dU\rangle
\eeqa
with the mass matrix
$\chi =\mbox{diag}(m_{\pi}^2,m_{\pi}^2, 2m_K^2-m_{\pi}^2)$
and
\beqa
d^4x \, {\tilde \Lagr}_{\partial^2}&=&{\tilde K}_3\langle
(U^{\dagger} dv U + dv)(
[U^{\dagger}\partial^{\lambda}dU - 
\left(\partial^{\lambda}dU\right)^{\dagger} U] 
U^{\dagger}dU\,\,\, U^{\dagger}\partial_{\lambda}U\nonumber\\%[2mm]
&& \qquad \qquad \qquad  \qquad 
+ U^{\dagger}\partial_{\lambda}U\,\,\, 
U^{\dagger}d U[U^{\dagger}\partial^{\lambda}dU - 
(\partial^{\lambda}dU)^{\dagger} U])
\rangle \nonumber\\%[2mm]
&+&{\tilde K}_4\langle
(U^{\dagger} dv U + dv)(
[U^{\dagger}\partial^{\lambda}d U - 
(\partial^{\lambda} d U)^{\dagger} U] 
U^{\dagger} d U\,\,\, U^{\dagger}\partial_{\lambda} U\nonumber\\%[2mm]
&& \qquad \qquad \qquad  \qquad 
+ U^{\dagger}\partial_{\lambda}U\,\,\, 
U^{\dagger}d U [U^{\dagger}\partial^{\lambda}d U - 
(\partial^{\lambda} d U)^{\dagger} U])
\rangle .
\eeqa
At next-to-leading order we replace the charge matrix $Q$ by $Q^{(0)}$,
since $Q^{(1)}$ contributes beyond our working precision.
Without mixing the counter terms yield the amplitudes
\beq  \label{nloamp}
{\cal A}^{(ct)}(\phi \to  \pi^+ \pi^- \gamma)= \frac{8 e}{\sqrt{3} f^3}
       k_\mu \epsilon_\nu p^+_\alpha p^-_\beta \epsilon^{\mu \nu \alpha \beta} 
       \beta_\phi 
\eeq
with the coefficients
\beqa
\beta_{\eta_8} &=&  m_\pi^2 \left[2{\tilde K}_1+{\tilde K}_2\right]  
                - [m_{\eta}^2+2 s_{+-}-2 m_{\pi}^2] {\tilde K}_3 -
                  [ s_{+-}-2 m_{\pi}^2] {\tilde K}_4 \no\\
\beta_{\eta_0} &=&  \frac{3}{\sqrt{2}} {\tilde L}_1
              +  \sqrt{2} m_\pi^2 \left[2{\tilde K}_1+{\tilde K}_2\right]
              - \sqrt{2} [m_{\eta'}^2+2 s_{+-}-2 m_{\pi}^2] {\tilde K}_3 -
                 \sqrt{2} [ s_{+-}-2 m_{\pi}^2] {\tilde K}_4
\eeqa
and $s_{+-}= (p^+ + p^- )^2$.
One must furthermore account for the $Z$-factors of the mesons and 
$\eta$-$\eta'$
mixing up to next-to-leading order.
For each pion leg the pertinent $Z$-factor
\beq
\sqrt{Z}_{\pi}=1-\frac{4}{f^2}m_{\pi}^2L_{5}^{(r)}
\eeq
can be completely absorbed by replacing one factor of $f$ by the physical decay
constant $F_{\pi}$ in the denominator of the amplitude, Eq.~(\ref{nloamp}),
\beq
F_{\pi}=f\left(1+\frac{4}{f^2}m_{\pi}^2 L_{5}^{(r)}\right) .
\eeq
The coupling constant $L_{5}^{(r)}$ originates from the 
effective Lagrangian of natural parity
\beq
{\cal L}_{\scriptscriptstyle{eff}} = {\cal L}^{(0)} + {\cal L}^{(1)}  + \ldots
\eeq
which reads at lowest order $\delta^0$
\beq
{\cal L}^{(0)} = \frac{f^2}{4} \langle \partial_\mu U ^\dagger 
\partial^\mu U \rangle 
+ \frac{f^2}{4} \langle \chi U ^\dagger +  U \chi^\dagger \rangle  - 
\frac{1}{2} \tau \psi^2 
\eeq
and at next-to-leading order ${\cal O}(\delta)$  
\beqa
{\cal L}^{(1)} &=& L_5 \langle \partial_\mu U ^\dagger \partial^\mu 
U (\chi^\dagger U + U^\dagger \chi) \rangle +
 L_8 \langle \chi^\dagger U \chi^\dagger U  + U^\dagger \chi U^\dagger \chi 
\rangle \no \\
&+& \frac{f^2}{12} \Lambda_1 \partial_\mu \psi \partial^\mu \psi + 
i \frac{f^2}{12} \Lambda_2 \psi 
\langle \chi^\dagger U - U^\dagger \chi \rangle .
\eeqa
Note that both $L_5$ and $L_8$ contain divergent pieces which compensate
divergencies from loop integrals at order ${\cal O}(\delta^2)$
and are thus suppressed by one order in $N_c$ with respect to the finite parts
$L_{5}^{(r)}, L_{8}^{(r)}$. 
To the order we are working, we omit the divergent portions.

In the tree level expression for the decay amplitude, Eq.~(\ref{treeamp}),
the states $\eta_8$ and $\eta_0$ are replaced by the physical states
$\eta$ and $\eta'$ via \cite{B}
\beqa \label{eq:mix}
\frac{1}{f}\eta_8&=&\frac{1}{F_\eta^8}
[\mbox{cos}\vartheta^{(1)} - \mbox{sin}\vartheta^{(0)}{\cal A}^{(1)}] \, \eta
+ \frac{1}{F_\eta^8}
[\mbox{sin}\vartheta^{(1)} +\mbox{cos}\vartheta^{(0)}{\cal A}^{(1)}] \, \eta'
\nonumber\\
\frac{1}{f}\eta_0&=&\frac{1}{F_{\eta'}^0}
[\mbox{cos}\vartheta^{(0)}{\cal A}^{(1)} - \mbox{sin}\vartheta^{(1)}] \, \eta
+ \frac{1}{F_{\eta'}^0}
[\mbox{sin}\vartheta^{(0)}{\cal A}^{(1)} +\mbox{cos}\vartheta^{(1)}] \, \eta'
  ,
\eeqa
where
\beqa
{\cal A}^{(1)} &=&\frac{8\sqrt{2}}{3 F_{\pi}^2}L_5^{(r)}[
m_{\scriptscriptstyle{K}} - m_{\pi}^2] ,
\nonumber\\
\mbox{sin}2\vartheta^{(1)} &=& \mbox{sin}2\vartheta^{(0)}\left(
\frac{1+\Lambda_2}{\sqrt{1+\Lambda_1}} 
+ \frac{8}{F_{\pi}^2}
[2 L_8^{(r)}-L_5^{(r)}](m_K^2-m_{\pi}^2)-
\frac{24}{F_{\pi}^4}L_5^{(r)}\tau
\right) .
\eeqa
The numerical discussion of these expressions is presented in \cite{B}.
For the counter term contributions in Eq.~(\ref{nloamp}), on the other hand,
we keep only the leading order pieces in Eq.~(\ref{eq:mix})
\beqa
\frac{1}{f}\eta^8&=&\frac{1}{F_\eta^8}
\mbox{cos}\vartheta^{(0)}\eta +\frac{1}{F_\eta^8}
\mbox{sin}\vartheta^{(0)}\eta' ,
\nonumber\\
\frac{1}{f}\eta^0&=& -\frac{1}{F_{\eta'}^0}
\mbox{sin}\vartheta^{(0)}\eta
+\frac{1}{F_{\eta'}^0}
\mbox{cos}\vartheta^{(0)}\eta'
\eeqa
which was already employed in the discussion of the leading order decay 
amplitude, cf. Eq.~(\ref{eq:leadmix}).

From our results it is easy to see that the $\eta'$ decay does not 
depend on the
QCD renormalization scale.
Due to the anomalous dimension of the singlet axial current, the decay constant
$F_{\eta'}^0$ scales as, cf. Eq.~(\ref{decconst}),
\beq
F_{\eta'}^0 \to Z_A F_{\eta'}^0 \; ,
\eeq
where $Z_A$ is the multiplicative renormalization constant of the 
singlet axial current.
Furthermore, the ${\tilde K}_i$ are scale independent, whereas ${\tilde L}_1$
transforms according to \cite{KL1}
\beq
{\tilde L}_1\to{\tilde L}_1^{ren}=Z_{\scriptscriptstyle{A}}{\tilde L}_1 
- \frac{N_{\scriptscriptstyle{C}}}{144\pi^2}[Z_A-1] .
\eeq
Since ${\tilde L}_1$ appears in the $\eta'$ decay amplitude
in the combination
\beq
\left(\frac{N_{\scriptscriptstyle{C}}}{12\pi^2}-12{\tilde L}_1\right)\to
\frac{N_{\scriptscriptstyle{C}}}{12\pi^2}-12{\tilde L}_1^{ren}=
Z_A\left(\frac{N_{\scriptscriptstyle{C}}}{12\pi^2}-12{\tilde L}_1\right) ,
\eeq
the amplitude remains renormalization group invariant.

\begin{figure}[t]
\begin{center}
\includegraphics[height=4.cm,clip]{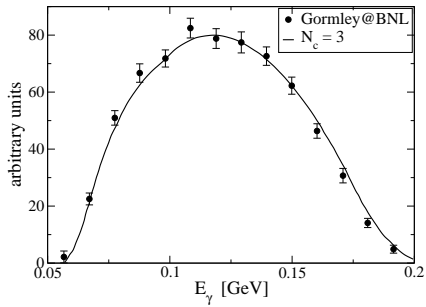}
\end{center}
\caption{Photon spectrum for $N_C=3$}
\label{NC3}
\end{figure}

\begin{figure}[t]
\begin{center}
\includegraphics[height=4.0cm,clip]{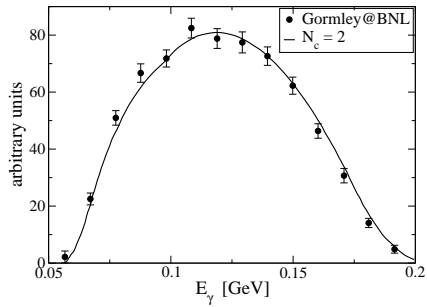}
\end{center}
\caption{Photon spectrum for $N_C=2$}
\label{NC2}
\end{figure}

We now determine the unknown coefficients ${\tilde K}_i$ 
by fitting them to both the decay width $\Gamma_{\eta \to \pi^+ \pi^- \gamma}$
and the corresponding photon spectrum.
To this end, we rewrite the coefficient $\beta_{\eta_8} $ 
in terms of effectively two parameters
\beqa
\beta_{\eta_8}  \equiv  \beta_{\eta_8} ^{(1)} +   
                \beta_{\eta_8} ^{(0)} s_{+-}  .
\eeqa
Setting $N_c=3$ we obtain a perfect fit to both the experimental decay width
$\Gamma_{\eta \to \pi^+ \pi^- \gamma} = 56.1 \pm 5.4$ eV and the photon 
spectrum,
see Fig.~\ref{NC3}
, with $\beta_{\eta_8} ^{(1)}=1.3\times10^{-3}$ and 
$\beta_{\eta_8} ^{(0)} =28.4\times10^{-3}\mbox{GeV}^{-2}$ 
which shows that the subleading
contributions from the counter terms are important and not suppressed with
respect to the leading order originating from the WZW term. 
However, for $N_c=2$ an equally good fit to the experimental data, see
Fig.~\ref{NC2},
is achieved by setting $\beta_{\eta_8} ^{(1)}=-3.2\times 10^{-3}$ and 
$\beta_{\eta_8} ^{(0)} =22.0\times10^{-3}\mbox{GeV}^{-2}$. 
Although a fit for $N_c=1$ would be possible as well, we do not present the 
results here, as a world with $N_c=1$ has no strong interactions.
Note that in the present work we do not explore the possibility of estimating 
the values of the unknown couplings by means of model-dependent assumptions 
such as resonance saturation.

It thus does not seem to be possible to strictly determine the number of colors
at next-to-leading order in large $N_c$ ChPT, unless one imposes in addition 
the cancellation of Witten's global $SU(2)_L$ anomaly which requires $N_c$ to 
be odd \cite{Wit}. In that case $N_c=2$ is ruled out and for $N_c=5$
it turns out that -- to the order we are working -- one cannot bring the 
results into agreement with experiment by varying the couplings. In
particular, the photon spectrum can only  be reproduced with a larger decay 
width. One may be inclined to argue that the restriction to odd $N_c$ enables 
a determination of $N_c$, but it is well-known from the one-loop calculation 
of this decay in conventional ChPT that the loop contributions reduce the 
decay width \cite{BN2, BBC}.
It is therefore possible that a next-to-next-to-leading order calculation
in large $N_c$ ChPT including one-loop corrections can be brought to agreement
with experiment also for $N_c=5$. However, such an investigation is beyond the
scope of the present work.
In any case, a rigorous statement on the number of colors cannot be made due 
to the
failure of the anomalous contribution from the WZW term to accomodate the
decay width for $N_c=3$ and the presence of unknown couplings.

In the case of the $\eta'$ decay unitarity effects via final state 
interactions 
are dominating \cite{CB,GAMS}. Therefore, a perturbative approach is
insufficient to describe
the $\eta'$ decay, and we will refrain from presenting numerical results here.

We conclude that a clean derivation of the number of colors cannot be achieved
by investigating the decays 
$\eta, \eta' \rightarrow \pi^+ \pi^- \gamma$. In particular, $\eta \rightarrow 
\pi^+ \pi^- \gamma$ should not be 
utilized as a textbook example to confirm the number of colors to be $N_c=3$.\\

{\it  We thank
Robin Ni{\ss}ler for useful discussions. Financial support of the DFG is
gratefully acknowledged.}

%%%%%%%%%%%%%%%%%%%%%%%%%%%%%%%%%%%%%%%%%%%%%%%%%%%%%%%%%%%%%%%%%%%%%%%%%%%%%%

\end{document}